\documentclass[cameraready]{Interspeech}

\title{UniSE: A Unified Framework for Decoder-Only Autoregressive LM-Based Speech Enhancement}

\author[affiliation={1}, equalcontribution]{Haoyin}{Yan}
\author[affiliation={1}, equalcontribution]{Chengwei}{Liu}
\author[affiliation={1,2}, correspondingauthor]{Shaofei}{Xue}
\author[affiliation={1}, ]{Xiaotao}{Liang}
\author[affiliation={1}, ]{Yinghao}{Liu}
\author[affiliation={2}, ]{Yuxiang}{Kong}
\author[affiliation={1}, ]{Zheng}{Xue}

\address{
    $^1$ Qwen Business Unit of Alibaba, China \\
    $^2$ Tongyi AI Lab of Alibaba, China
}

\email{yanhaoyin.yhy@alibaba-inc.com, liuchengwei.lcw@alibaba-inc.com}


\usepackage{comment}

\usepackage{cite}
\usepackage{amsmath,amssymb,amsfonts}
\usepackage{algorithmic}
\usepackage{graphicx}
\usepackage{textcomp}
\usepackage{xcolor}

\usepackage{url}
\usepackage{color}
\usepackage{bm}
\usepackage{booktabs}
\usepackage{threeparttable}
\usepackage{verbatim}
\usepackage{multirow}
\usepackage{setspace}
\usepackage{etoolbox}
\usepackage{float}

\begin{document}

\maketitle

\begin{abstract}
Neural audio codecs have largely promoted the application of language models (LMs) for speech applications.
However, the effectiveness of autoregressive LM-based models in unifying speech enhancement (SE) tasks remains underexplored.
In this work, we propose UniSE, a unified decoder-only LM-based framework to handle different SE tasks including speech restoration, 
target speaker extraction, and speech separation.
Conditioned on input speech features, 
it autoregressively generates target discrete tokens, 
facilitating compatibility between distinct learning patterns of multiple tasks. 
To further optimize speech quality, 
we introduce a progressive reinforcement learning strategy 
with multiple assessment criteria.
Experiments on several benchmarks show that UniSE achieves competitive performance compared to discriminative and generative baselines, 
demonstrating the capacity of LMs in unifying SE tasks.
The code and demo are available at: https://github.com/alibaba/unified-audio/tree/main/QuarkAudio-UniSE.
\end{abstract}

\section{Introduction}
\label{sec:intro}

Recently, 
speech enhancement (SE) has expanded beyond conventional denoising
toward reconstructing
clean target speech from degraded recordings~\cite{VoiceFixer, use}. 
In this context, SE can include many sub-tasks: speech restoration (SR) that
aims to restore speech from the degraded recordings with various distortions; target speaker extraction (TSE) that extracts the target speech guided by an assistive clue, e.g., reference speech of the target speaker; 
speech separation (SS) that aims at separating all existing speakers from the mixture. 
Deep neural networks achieve better performance in non-stationary scenarios than traditional algorithms
and thus become the mainstream in this field~\cite{ConvTasNet,CMGAN}.

Language models (LMs) have achieved remarkable success in generating 
text~\cite{qwen2}, 
images~\cite{VAR} and 
audio~\cite{Audiobox,SparkTTS}.
Some works have explored applications of LMs 
to SE by typically predicting the discrete tokens of clean speech, which are extracted by pre-trained neural audio codecs (NACs).
For instance, 
GenSE~\cite{GenSE} is a two-stage approach based on autoregressive (AR) modeling, 
where the first stage generates clean semantic tokens in noisy semantic conditions, 
and the outputs are utilized to predict clean acoustic features in the second stage. 
In~\cite{LauraTSE}, a TSE model called LauraTSE 
extracts continuous features of 
mixture and reference speech, serving as prefixes to estimate the target discrete tokens. 
Although these works have shown the potential 
of LMs in SE, they are confined to single distortion or task, 
resulting in limited extensibility to diverse scenarios.

Some studies consider more distortions or focus on the unification of multiple tasks to expand the universality of SE systems.
MaskSR~\cite{MaskSR} handles additive noise, reverberation, clipping and bandwidth limitation
via masked prediction~\cite{MaskGIT} on multi-layer discrete tokens.
LLaSE-G1~\cite{LLaSEG1} employs a non-autoregressive (NAR) LM to map noisy WavLM~\cite{WavLM} features to the clean discrete tokens, 
with a dual-channel input and output architecture to support multiple tasks.
These works involve the paradigm of masked generation or direct mapping, 
and the effectiveness of AR modeling in multi-task SE frameworks remains to be further verified. 
Considering the flexible prefix formulations of the decoder-only model, 
it has potential to act as an elegant solution for the task unification. 

In the context of Large Language Models (LLMs), 
reinforcement learning (RL) has been primarily employed to align 
model behavior with human preferences and task-specific objectives~\cite{rl_survey}.
Some works have verified the potential of RL in the domain of text-to-speech (TTS). 
For instance, the Direct Preference Optimization (DPO)~\cite{dpo} framework is utilized to improve the emotional consistency of synthesized speech in~\cite{emo_dpo}.
However, the application of RL in SE remains largely unexplored. 
While traditional SE systems rely on supervised learning with signal-level losses (e.g., SI-SNR), 
these metrics do not always correlate well with auditory perception. 
This gap motivates the exploration of RL-based alignment in SE: 
by leveraging perceptual rewards as feedback, 
the RL has potential to guide the SE model towards better perceptual alignment.

\begin{figure*}
    \centering
    \includegraphics[width=0.85\textwidth]{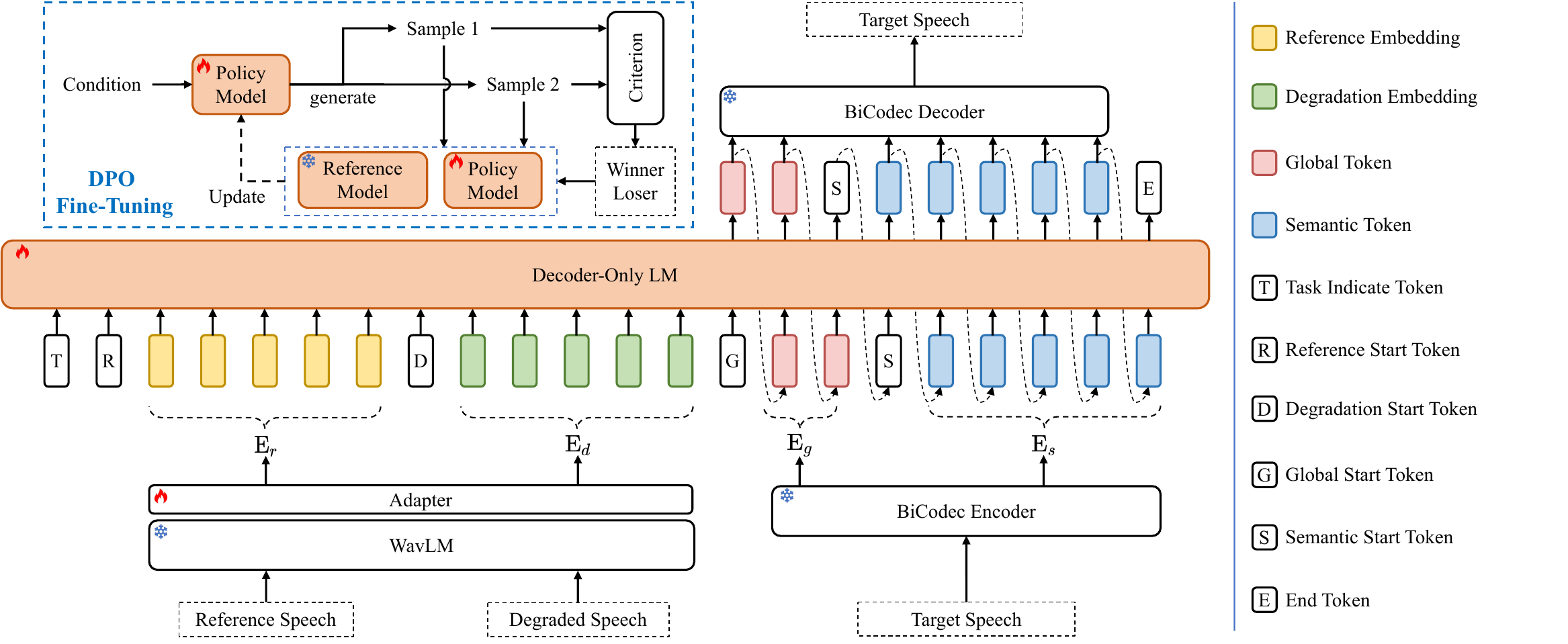}
    \caption{Overall architecture of UniSE, where the BiCodec Encoder is only utilized to generate label tokens during training and excluded during inference.
    The snowflake icon means that parameters are pre-trained and frozen, and the fire icon indicates that  parameters are optimized during training.
    }
    \label{fig:architecture}
    \vspace{-0.6em}
\end{figure*}

In this work, we propose a decoder-only AR LM-based framework called \textbf{UniSE} to unify multiple sub-tasks of SE, including SR, 
TSE and SS. Our contributions are fourfold:
1) We design a decoder-only SE framework, which utilizes continuous conditional features to 
generate discrete tokens of target speech.
2) We propose a task token to distinguish between different operational modes, unifying multiple tasks by switching and combination of these modes. 
3) We introduce a progressive 
reinforcement learning (PRL) strategy to fine-tune our model, 
where criteria (i.e., perceptual quality and similarity) are progressively introduced across different stages
to perform DPO optimization.
4) Our model achieves competitive performance on several benchmarks, revealing the potential of decoder-only AR LM in the unification of SE sub-tasks.

\section{METHODOLOGY}
\label{sec:metho}

Fig.~\ref{fig:architecture} illustrates the UniSE framework, comprising: (1) a pre-trained WavLM model with a trainable adapter for continuous speech feature extraction; (2) a discrete speech codec for tokenization and waveform reconstruction; (3) a decoder-only language model (LM) backbone for autoregressive (AR) conditional modeling; and (4) a DPO fine-tuning framework for perceptual quality enhancement.

\subsection{Conditional Feature Extractor}
To extract conditioning features from reference and degraded speech input, we adopt the pre-trained WavLM\footnote{https://huggingface.co/microsoft/wavlm-base-plus} as the feature extractor.
We average the features from all transformer layers in WavLM to obtain sufficient acoustic and semantic information simultaneously.
A trainable linear adapter maps the output from frozen WavLM into a representation space amenable to LM AR modeling, 
yielding features ${\rm E}_r$ and ${\rm E}_d$ for reference and degraded speech, respectively.

\subsection{Discrete Token Codec}
We utilize BiCodec~\cite{SparkTTS} to convert the continuous regression problems of SE into discrete autoregressive modeling.
During training, the BiCodec Encoder produces a fixed-length global feature ${\rm E}_g$ (32 tokens, speaker characteristics) and a variable-length semantic feature ${\rm E}_s$ (50 tokens/s, speech content), 
both with single-layer quantization for easy AR integration.
During inference, the BiCodec decoder reconstructs the original speech by combining ${\rm E}_g$ and ${\rm E}_s$, leveraging this explicit disentanglement for high-fidelity restoration.
which benefits from this explicit disentanglement to achieve high fidelity.

\subsection{Unified Multi-Task Framework}
The proposed framework adopts the LLaMA architecture~\cite{LLaMA} as the backbone for AR modeling, aiming to estimate the conditional probability distribution of target speech discrete representations given optional reference and degraded speech inputs. 
To unify SR, TSE and SS tasks within a single framework, we define three operational modes: SR mode, TSE mode and reverse TSE (rTSE) mode. Each mode corresponds to a learnable task-specific token: ${\rm T_{SR}}$, ${\rm T_{TSE}}$, and ${\rm T_{rTSE}}$.
In SR mode, the target speech corresponds to the clean signal of the degraded input. The input sequence of AR LM is formatted as 
[${\rm T_{SR}}$, ${\rm D}$, ${\rm E}_d$, ${\rm G}$, ${\rm E}_g$, ${\rm S}$, ${\rm E}_s$], 
where ${\rm D}$ denotes the start of degraded speech features, ${\rm G}$ the start of global features, and ${\rm S}$ the start of semantic features, 
respectively. The output sequence is formulated as ${\bm o}= \left[ {\rm E}_g, {\rm S}, {\rm E}_s, {\rm E} \right]$, with ${\rm E}$ representing 
the end-of-sequence token. The parameters $\theta$ of the adapter and decoder-only LM are optimized by minimizing the negative log-likelihood of the predicted outputs:
\begin{align}
\mathcal{L}_{\rm SR} = - \sum_{t=1}^{L} {\rm log}P\left(o_t | {\rm T_{SR}}, {\rm D}, {\rm E}_d, o_{<t} ; \theta\right),
\end{align}
where $L$ indicates the length of output sequence. 

For the TSE mode, the target speech corresponds to the timbre-matched speech component 
in the degraded input that aligns with the reference audio. The input sequence is formatted as 
[${\rm T_{TSE}}$, ${\rm R}$, ${\rm E}_r$, ${\rm D}$, ${\rm E}_d$, ${\rm G}$, ${\rm E}_g$, ${\rm S}$, ${\rm E}_s$], 
where ${\rm R}$ denotes the start of reference speech features. The associated loss function is defined as
\begin{align}
\mathcal{L}_{\rm TSE} = - \sum_{t=1}^{L} {\rm log}P\left(o_t | {\rm T_{TSE}}, {\rm R}, {\rm E}_r, {\rm D}, {\rm E}_d, o_{<t} ; \theta\right).
\end{align}
For the rTSE mode, the target speech corresponds to the timbre-mismatched 
component in the degraded input when compared with 
the reference audio. The input sequence format and loss function $\mathcal{L}_{\rm rTSE}$ keep identical to that of the TSE mode.

\subsection{Progressive Reinforcement Learning (PRL)}
After the training stage of UniSE, we fine-tune it using DPO~\cite{dpo} framework to further improve the 
perceptual quality of the generated speech, 
as illustrated in Fig.~\ref{fig:architecture}. 
The pre-trained UniSE 
serves as the reference model and initializes the policy model.
During fine-tuning, the policy model generates two candidate results on-the-fly when given the condition $c$ of 
reference and degraded speech. 
After determining the winner sample $y_w$ and loser sample $y_l$ using specific criterion,
the DPO loss is calculated as
{\footnotesize
\begin{align}
\mathcal{L}_{\rm DPO} = - \mathbb{E} \log \sigma \left[ \beta \left( \log\frac{\pi_\theta(y_w | c)}{\pi_{r}(y_w | c)} - \log\frac{\pi_\theta(y_l | c)}{\pi_{r}(y_l | c)} \right) \right],
\end{align}
}
where $\pi_\theta$ and $\pi_{r}$ denote the
sequence-level probabilities under the current policy model and reference model, 
$\beta > 0$ is a temperature hyperparameter controlling the strength of the preference signal, 
and $\sigma$ indicates the sigmoid function. 

To better align the model with human subjective perception, we divide the fine-tuning into two progressive stages.
In stage 1 (S1), we adopt DNSMOS~\cite{DNSMOS} as the preference criterion,
which predicts three sub-scores: SIG (signal fidelity), BAK (background noise level), and OVRL (overall quality).
The average of these three scores is used to determine the winner and loser samples,
and the fine-tuning loss is formulated as
\begin{align}
\mathcal{L}_{\rm S1} = \alpha \mathcal{L}_{\rm CE} + \mathcal{L}_{\rm DPO}^{\rm DNSMOS},
\label{eq:stage1}
\end{align}
where $\mathcal{L}_{\rm CE}$ denotes the cross-entropy loss used during pre-training,
$\alpha \geq 0$ controls its contribution,
and $\mathcal{L}_{\rm DPO}^{\rm DNSMOS}$ is the DPO loss with preference determined by DNSMOS.
In stage 2 (S2), we additionally introduce 
the similarity of the WavLM features
as the criterion to optimize semantic and acoustic consistency.
Specifically, the WavLM features of candidate samples and the target speech are computed,
and the Euclidean distance is 
computed to determine the winner and loser samples, 
where smaller distance indicates higher similarity to the ground truth.
The fine-tuning loss in this stage is defined as
\begin{align}
\mathcal{L}_{\rm S2} = \alpha \mathcal{L}_{\rm CE} + 0.5 \mathcal{L}_{\rm DPO}^{\rm DNSMOS} + 0.5 \mathcal{L}_{\rm DPO}^{\rm WavLM},
\label{eq:stage2}
\end{align}
where $\mathcal{L}_{\rm DPO}^{\rm WavLM}$ is the DPO loss with preference determined by the WavLM feature distance. 
Gradually increasing the complexity of preference signals 
allows the model to first establish a stable foundation of perceptual quality 
before refining details, 
thereby avoiding conflicting optimization objectives.

\subsection{Inference Strategies}
To alleviate computational overhead during inference, input speech is segmented into fixed-length chunks consistent with the training configuration, as described in MaskSR~\cite{MaskSR}. The SR mode is utilized for SR task, which restores clean speech from the degraded recording. 
When multiple speakers exist in the degraded speech, our model intends to output the louder speaker. 
The TSE mode processes the TSE task, which extracts timbre-matched speech from the mixture regardless of the relative loudness.
While for the SS task, we consider multiple inferences that involve all three modes. 
Specifically, for two-speaker SS (this work only considers the two-speaker case),
we first employ
the SR mode to extract the louder speaker. This initial result then serves as the reference for the TSE mode to 
separate the first speaker, which ensures speaker consistency across different segments when relative loudness varies.
Finally, the rTSE mode  is applied to extract the remaining speaker, using the first speaker as a reference.

\begin{table}[t]
    \centering
    \caption{Distortion categories and simulation configurations, where SNR denotes signal-to-noise ratio
    and SIR denotes signal-to-interference ratio.}
    \vspace{-1em}
    \resizebox{\columnwidth}{!}{
        \begin{tabular}{lcc}
            \toprule
            Distortion & Probability & Hyperparameters  \\
            \midrule
            \midrule
            Noise & 0.8 & SNR $\in$ [-5, 20] \\
            \midrule
            Reverberation & 0.3 & - \\
            \midrule
            \multirow{2}{*}{Clipping} & \multirow{2}{*}{0.3} & Min\_quantile $\in$ [0.0, 0.1] \\
            &  & Max\_quantile $\in$ [0.9, 1.0] \\
            \midrule
            Bandwidth Limitation & 0.3 & Bandwidth $\in$ \{2, 4\} kHz \\
            \midrule
            Packet Loss & 0.3 & Rate $\in$ [0.05, 0.25] \\
            \midrule
            \multirow{2}{*}{Interference Speaker} & 0.2 for SR & SIR $\in$ [2, 20] for SR \\
            & 1.0 for TSE/rTSE & SIR $\in$ [-5, 5] for TSE/rTSE \\
            \bottomrule
        \end{tabular}
    }
    \vspace{-2em}
    \label{tab:simu}
\end{table}

\section{EXPERIMENTS}
\label{sec:exp}

\subsection{Experimental Setup}

\textbf{Training Datasets:} The clean speech data for training is sourced from the VoxBox dataset~\cite{SparkTTS}, an integrated and rigorously cleaned compilation of multiple public datasets.
This set comprises 760 hours of LibriSpeech~\cite{Librispeech} data, 1200 hours from the MLS\_English~\cite{MLS} subset, 
and 1800 hours of the Emilia\_ZH~\cite{Emilia} subset. The noise corpus comprises approximately 460 hours of data from the DNS Challenge~\cite{DNS}, 
FSD50K~\cite{FSD50K}, WHAM!~\cite{WHAM}, DESED~\cite{DESED}, DEMAND~\cite{DEMAND}, MUSAN~\cite{MUSAN}, DISCO~\cite{DISCO}, MUSDB18-HQ~\cite{MUSDB18HQ}, and TUT Urban Acoustic Scenes~\cite{UAS}.
Additionally, we include 60,000 room impulse response (RIR) samples from SLR28 to simulate reverberation. 
A data augmentation pipeline simulates degraded speech, detailed in Table~\ref{tab:simu}.
We randomly select operational modes during training, 
and distortions are applied based on the given probability.
All audio samples are sampled to 16 kHz.

\noindent \textbf{Implementation Details:} The LLaMA-based decoder-only backbone consists of 12 layers, 
each with 8 attention heads and a hidden dimension of 512, resulting in 63M parameters. 
Our model is trained using AdamW optimizer with 30 epochs. 
The learning rate reaches a peak of 1e-3 after 4000 warm-up steps, then decays by a factor of 0.98 per epoch. For fine-tuning, the factor $\alpha$ is set to 0.4 and the model is updated for 5k steps with a batch size of 32
and a learning rate of 5e-5.
During training and inference, the lengths of reference and degraded speech are clipped/padded to 5 seconds.

\begin{table}[t]
    \centering
    \caption{DNSMOS scores on DNS 2020 Challenge test sets, where ``With Reverb'' subset contains reverberation while 
    ``No Reverb'' subset only involves noise. 
    }
    \vspace{-1em}
    \resizebox{\columnwidth}{!}{
        \begin{tabular}{lcccccccc}
            \toprule
            \multirow{2}{*}{Model} & \multirow{2}{*}{Para. (M)} & \multirow{2}{*}{MACs (G/s)} &  \multicolumn{3}{c}{With Reverb} & \multicolumn{3}{c}{No Reverb} \\
            \cmidrule(lr){4-6} \cmidrule(lr){7-9}
            & & & SIG &	BAK &	OVRL & SIG & BAK & OVRL \\
            \midrule
            \midrule
            Noisy &	-	& - & 1.76 & 1.50 & 1.39 & 3.39 & 2.62 & 2.48 \\
            \midrule
            \multicolumn{9}{c}{\textit{Discriminative Methods}} \\
            \midrule
            Conv-TasNet~\cite{ConvTasNet} & 5.1 & 5.2 & 2.42 & 2.71 & 2.01 & 3.09 & 3.34 & 3.00 \\
            FRCRN~\cite{frcrn} & 10.3 & 12.3 & 2.93 & 2.92 & 2.28 & 3.58 & 4.13 & 3.34 \\
            TF-GridNet~\cite{tf_gridnet} & 2.8 & 49.6 & 3.04 & 3.66 & 2.63 & 3.58 & 4.17 & 3.35 \\
            \midrule
            \multicolumn{9}{c}{\textit{Generative Methods}} \\
            \midrule
            SELM~\cite{SELM} & - & - & 3.16 & 3.58 & 2.70 & 3.51 & 4.10 & 3.26 \\
            MaskSR~\cite{MaskSR} & - & - & 3.53 & 4.07 & 3.25 & 3.59 & 4.12 & 3.34 \\
            AnyEnhance~\cite{AnyEnhance} & 363.5 & - & 3.50 & 4.04 & 3.20 & 3.64 & 4.18 & 3.42 \\
            GenSE~\cite{GenSE} & 667.6 & 99.0 & 3.49 & 3.73 & 3.19 & 3.65 & 4.18 & 3.43 \\
            LLaSE-G1~\cite{LLaSEG1} & 1895.6 & 63.9 & 3.59 & 4.10 & 3.33 & 3.66 & 4.17 & 3.42 \\
            \midrule
            UniSE & 263.9 & 42.2 & 3.68 & 4.12 & 3.41 & 3.67 & 4.13 & 3.42 \\
            UniSE-SR & 263.9 & 42.2 & 3.66 & 4.08 & 3.38 & 3.66 & 4.14 & 3.42 \\
            UniSE + S1 & 263.9 & 42.2 & 3.77 & 4.21 & 3.55 & 3.73 & 4.19 & 3.51 \\
            UniSE + S2 & 263.9 & 42.2 & 3.78 &	4.23 & 3.57 & 3.74 & 4.20 & 3.53 \\
            UniSE + PRL & 263.9 & 42.2 & \textbf{3.83} & \textbf{4.26} & \textbf{3.64} & \textbf{3.76} & \textbf{4.23} & \textbf{3.57} \\
            \bottomrule
        \end{tabular}
    }
    \vspace{-0.5em}
    \label{tab:sr1}
\end{table}


\begin{table}[t]
    \centering
    \caption{SR results on URGENT 2025 Challenge blind test set.}
    \vspace{-1em}
    \footnotesize
    \resizebox{\columnwidth}{!}{
        \begin{threeparttable}
        \begin{tabular}{lcccc}
            \toprule
            Team/Model & Team Rank\tnote{\textdagger} & OVRL & NISQA & UTMOS \\
            \midrule
            \midrule
            Bobbsun &	1 & 2.88 & 3.22 & 2.09 \\
            Xiaobin (TF-GridNet) & 2 & 2.92 & 3.24 & 2.16 \\
            subatomicseer & 3 & 2.94 & 3.25 & 2.19 \\
            wataru9871 & 13 & 3.10 & 3.74 & 2.53 \\
            \midrule
            LLaSE-G1~\cite{LLaSEG1} & - & 2.80 & 2.93 & 2.09 \\
            \midrule
            UniSE & - & 3.17 & 3.68 & 2.85 \\
            UniSE + PRL & - & \textbf{3.34} & \textbf{3.83} & \textbf{2.95} \\
            \bottomrule
        \end{tabular}

        \begin{tablenotes}
        \footnotesize
        \item[\textdagger] The ranking takes into account both non-intrusive and intrusive metrics, where the 
        latter are not friendly to generative models.
        \end{tablenotes}

        \end{threeparttable}
    }
    \vspace{-2em}
    \label{tab:sr2}
\end{table}

\begin{table}[t]
    \centering
    \caption{TSE results on Libri2Mix clean test set.}
    \vspace{-1em}
    \resizebox{\columnwidth}{!}{
        \begin{tabular}{lccccccc}
            \toprule
            Model & Type & SIG & BAK & OVRL & NISQA & SIM \\
            \midrule
            \midrule
            Mixture & - & 3.38 & 3.10 & 2.65 & 2.45 & 0.85 \\
            \midrule
            Spex+~\cite{spex} & D & 3.38 & 3.77 & 3.00 & 3.03 & 0.96 \\
            WeSep~\cite{WeSep} & D & 3.56 & 3.93 & 3.23 & 4.04 & \textbf{0.99} \\
            TSELM-L~\cite{TSELM} & G & 3.55 & 4.08 & 3.23 & 4.03 & 0.91 \\
            AnyEnhance~\cite{AnyEnhance} & G & 3.64 & 4.07 & 3.35 & 4.28 & 0.91 \\
            LLaSE-G1~\cite{LLaSEG1} & G & 3.53 & 4.01 & 3.22 & 3.89 & 0.92 \\
            LauraTSE~\cite{LauraTSE} & G & 3.61 & 4.08 & 3.34 & \textbf{4.33} & 0.97 \\
            \midrule
            UniSE & G & 3.62 & 4.06 & 3.33 & 4.01 & 0.95 \\
            UniSE-TSE & G & 3.62 & 4.07 & 3.33 & 4.00 & 0.95 \\
            UniSE + PRL & G & \textbf{3.70} & \textbf{4.14} & \textbf{3.45} & 4.10 & 0.95 \\
            \midrule
            BiCodec & - & 3.59 & 4.05 & 3.30 & 4.02 & 0.97 \\
            \bottomrule
        \end{tabular}
    }
    \vspace{-0.75em}
    \label{tab:tse}
\end{table}

\begin{table}[t]
    \centering
    \caption{SS results on Libri2Mix and WSJ0-2mix test sets.}
    \vspace{-1em}
    \resizebox{\columnwidth}{!}{
        \begin{tabular}{lccccccc}
            \toprule
            \multirow{2}{*}{Model} & \multirow{2}{*}{Type} & \multicolumn{3}{c}{Libri2Mix} & \multicolumn{3}{c}{WSJ0-2mix} \\
            \cmidrule(lr){3-5} \cmidrule(lr){6-8}
            & & SIG &	BAK &	OVRL & SIG & BAK & OVRL \\
            \midrule
            \midrule
            Mixture & - & 2.33 & 1.66 & 1.64 & 3.42 & 3.20 & 2.76 \\
            \midrule
            Sepformer~\cite{Sepformer} & D & 3.33 & 3.88 & 3.02 & 3.43 & 3.96 & 3.14 \\
            Mossformer2~\cite{MossFormer2} & D & 3.44 & 3.94 & 3.11 & 3.50 & 4.05 & 3.23 \\
            LLaSE-G1~\cite{LLaSEG1} & G & 3.48 & 3.83 & 3.11 & 3.52 & 3.92 & 3.19 \\
            \midrule
            UniSE & G & 3.62 & 4.09 & 3.34 & 3.63 & 4.09 & 3.36 \\
            UniSE + PRL & G & \textbf{3.76} & \textbf{4.20} & \textbf{3.55} & \textbf{3.71} & \textbf{4.19} & \textbf{3.49} \\
            \bottomrule
        \end{tabular}
    }
    \vspace{-2em}
    \label{tab:ss}
\end{table}

\begin{table}[t]
   \centering
   \caption{Ablation study on DNS 2020 Challenge test sets.}
    \vspace{-1em}
   \footnotesize
   \resizebox{\columnwidth}{!}{
      \begin{tabular}{lcccccc}
        \toprule
        \multirow{2}{*}{Method} & \multicolumn{3}{c}{With Reverb} & \multicolumn{3}{c}{No Reverb} \\
        \cmidrule(lr){2-4} \cmidrule(lr){5-7}
        & SIG &	BAK &	OVRL & SIG & BAK & OVRL \\
        \midrule
        \midrule
        UniSE & \textbf{3.68} & \textbf{4.12} & \textbf{3.41} & \textbf{3.67} & 4.13 & 3.42 \\
        \midrule
        
        NAR & 3.39 & 3.70 & 2.96 & 3.62 & \textbf{4.14} & 3.39 \\
        Qwen2 & 3.67 & 4.10 & 3.40 & \textbf{3.67} & 4.13 & \textbf{3.43} \\
        X-codec2 & 3.57 & 4.03 & 3.27 & 3.60 & 4.09 & 3.34 \\
         \bottomrule
      \end{tabular}
   }
   \vspace{-0.75em}
   \label{tab:ablation}
\end{table}

\begin{figure}[t]
   \centerline{\includegraphics[width=0.8\columnwidth]{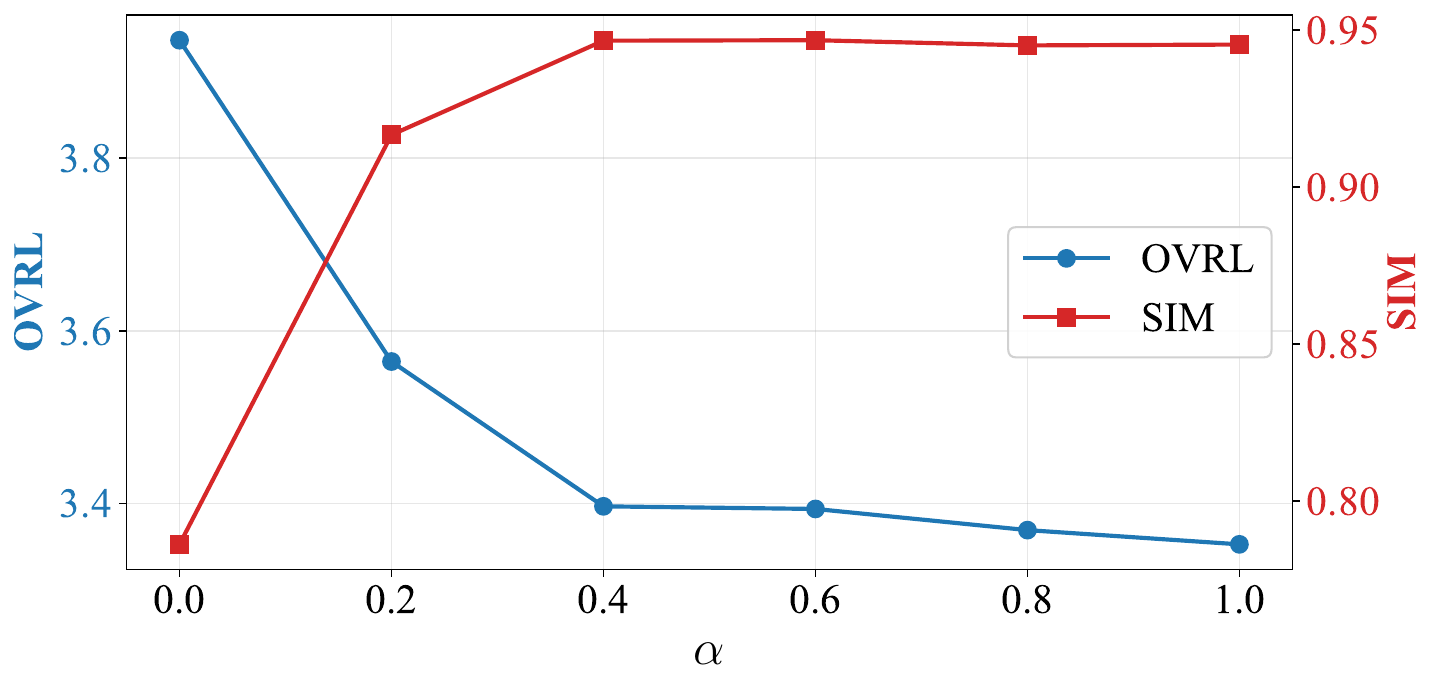}}
   \vspace{-0.5em}
   \caption{The OVRL and SIM scores in terms of different $\alpha$ values in the TSE task on 
   Libri2Mix clean test set.}
   \vspace{-2em}
   \label{fig:alpha}
\end{figure}

\noindent \textbf{Evaluation Configurations:} We evaluate our model on several benchmarks, including 
test sets from DNS 2020 Challenge~\cite{DNS}
and URGENT 2025 Challenge~\cite{URGENT} for SR task, Libri2Mix clean test set for the 
TSE task, and Libri2Mix noisy test set with WSJ0-2mix test set for the SS task.
We adopt DNSMOS~\cite{DNSMOS}, 
NISQA~\cite{NISQA} and UTMOS~\cite{UTMOS} to measure the quality of the generated speech. 
Following~\cite{LauraTSE}, the speaker similarity (SIM) 
is calculated using WavLM-base\footnote{https://huggingface.co/microsoft/wavlm-base-plus-sv} for TSE.

\subsection{Performance Comparison on Multiple Tasks}

Table \ref{tab:sr1} compares UniSE with advanced baselines on the DNS 2020 Challenge test set. Generative methods generally surpass discriminative counterpart, highlighting their potential for improving subjective listening quality. 
UniSE achieves state-of-the-art (SOTA) SR performance, 
and training exclusively in the SR mode (denoted as UniSE-SR) shows comparable performance with UniSE. 
Progressively introducing criteria outperforms 
optimization with single stage, 
indicating the coarse-to-fine strategy 
mitigates potential conflicts between heterogeneous preference signals 
and can lead to more stable convergence.
Table \ref{tab:sr2} evaluates our framework against URGENT Challenge submissions on a blind test set containing multiple distortions.
UniSE achieves competitive performance even under unseen distortions (codec artifacts, wind noise), demonstrating robust generalization ability.

TSE results on the  Libri2Mix clean test set are summarized in Table \ref{tab:tse}, 
showing that UniSE achieves performance comparable to SOTA baselines. UniSE supports a wider range of tasks than LauraTSE, an AR-based method with similar architecture and scale.
The UniSE-TSE variant trained exclusively in TSE mode 
achieves performance comparable to UniSE, mirroring the relationship between UniSE and UniSE-SR in the SR setting. This indicates that multi-task learning does not degrade individual task performance within our framework.
Fine-tuning with PRL strategy consistently enhances perceptual quality, as evidenced by improvements in DNSMOS and NISQA metrics.
Results produced by directly processing target speech using BiCodec (the bottom row) reveal the performance limitations 
of codecs on SE frameworks, demonstrating the necessity of further improving low-bitrate NACs.

Table \ref{tab:ss} compares the SS performance of our model with baselines on Libri2Mix and WSJ0-2mix test sets.
UniSE outperforms other discriminative and generative models with OVRL 
scores of 3.34 on Libri2Mix and 3.36 on WSJ0-2mix. 
The PRL fine-tuning further improves the score by 0.21 and 0.13 respectively.
This highlights the effectiveness of our multi-mode inference strategy and PRL fine-tuning.

\subsection{Ablation Studies}

We conduct ablation studies to evaluate the impact of modeling paradigms and architectures on the SE task.
Framing SE task as a mapping from degraded to target speech,we utilize NAR modeling (``\texttt{NAR}'') for semantic token prediction. Enhanced speech is then reconstructed from these tokens and ground-truth global tokens via the BiCodec decoder. However, the performance degrades notably, highlighting the advantage of AR modeling in capturing the underlying probability distribution of the data.

Moreover, replacing the LM backbone with Qwen2~\cite{qwen2} yields comparable results, demonstrating our framework's flexibility.
Conversely, utilizing X-codec2~\cite{xcodec2} leads to a clear performance decay, 
which can be attributed to its large codebook size that challenges the ability of LM. We also examine the weighting factor $\alpha$ values in~\eqref{eq:stage1} during S1 stage, as illustrated in Fig.~\ref{fig:alpha}.
When $\alpha$ is too low, 
the optimization is dominated by the DPO loss, 
leading to very high OVRL but a severe degradation in SIM, 
which reflects a mode collapse. 
Therefore, we choose $\alpha = 0.4$ to balance perceptual quality and speaker similarity.

\section{CONCLUSION}
In this work, we proposed UniSE, a unified framework for SE that integrates SR, TSE, and SS tasks. 
UniSE leverages continuous speech features as conditions to generate discrete target tokens via AR modeling. 
The use of task-specific tokens enables seamless switching and combination of multiple operational modes within a single model.
We further introduced a PRL fine-tuning strategy for coarse-to-fine optimization, aligning model generation with human perception. 
Extensive results show that our UniSE achieves competitive performance within each benchmark, verifying the effectiveness of decoder-only AR LM 
framework in unifying SE tasks. 
However, the proposed method faces limitations in strict streaming due to full-utterance conditioning and inferior decoding efficiency relative to NAR, posing worthy directions for future improvement.

\bibliographystyle{IEEEtran}
\bibliography{refs}

\end{document}